\documentclass[prb,twocolumn,showpacs,amssymb]{revtex4-1} 

\usepackage{amsmath, amssymb}
\usepackage[dvips]{graphicx}

\begin{document}
\title{Strong spintronic magnetoelectric effect in layered magnetic metamaterial}
\author{P.V.Pyshkin}
\author{A.V.Yanovsky}
\affiliation{B.Verkin Institute for Low Temperature Physics \&
Engineering,  National Academy of Sciences of Ukraine, 47 Lenin
Ave, 61103, Kharkov, Ukraine}

\begin{abstract}
It is shown that external magnetic field or magnetization induces electric polarization of microscopic isolated magnetic/non-magnetic hybrid structures due to the spin-dependent electron redistribution and mutual capacity. The magnetoelectric effect can be very strong, for instance the change of electric polarization reaches  $100$ $\mu C/m^{2}$ at a constant magnetic field of only $1.5$ Tesla in the case of $100$ $\div$ $10$ nanometer structures which tightly fill the space, composing a metamaterial. The effect does not require using exotic compounds. Based on the obtained results we suggest a number of recommendations for future experimental design.
\end{abstract}
\maketitle

Spin electronics (spintronics) has been already used in data storage and other devices. Advantages and physical principles of utilizing the spin of the charge carriers in electronics are widely discussed, see e.g. recent reviews \cite{ref1,ref1a} and the Nobel lectures of the 2007 \cite{ref2}. In this article we show that novel spintronic materials could manifest unique properties due to proximity between small magnetic/non-magnetic hybrid structures.

The magnetic/non-magnetic hybrid structures\cite{ref2a} are important subject of study in spintronics. Here the word ``non-magnetic'' refers to non-magnetic metal or semiconductor, such as Cu or GaAs. And the term ``magnetic'' means magnetic conductors, e.g. the ferromagnetic metals (Fe, Co), or some paramagnetic metals which are close to ferromagnetic instability (Pd), or the diluted magnetic semiconductors (Ga$_{1-x}$Mn$_x$As, Zn$_{1-x}$Be$_x$Se) etc. It is well known \cite{ref3} that magnetization induces the giant Zeeman spin splitting of the conduction band in such magnetic conductors due to the strong exchange interaction depending on the electron spin direction versus the magnetization axis. As is the convention, the appropriate spin bands are denoted hereinafter by the indices $\uparrow$ for spin-up or $\downarrow$  for spin-down.

The giant Zeeman splitting significantly changes the spin density of states (DOS) of the conduction electrons. It affects all the spin components of the physical parameters related to the DOS resulting giant magnetic resistance effect\cite{ref2, ref4}, spin-aligning\cite{ref3a} etc. Lots of magnetically inhomogeneous spintronic devices\cite{ref1} are proposed as parts of an electrical circuit: the spin-valve \cite{ref5}, Datta and Dass transistor \cite{ref6,ref7}, and other spin transistors \cite{ref8,ref9,ref10,ref11}.

In contrast, we discuss the equilibrium states of isolated conducting magnetic/non-magnetic structures after spin relaxation (i) with and (ii) without external magnetic field and show that heterogeneity of spin $\uparrow,\downarrow$ bands could be used to design artificial materials that display strong magnetoelectric (ME) effect.

We discuss direct ME effect: the induction of an electric polarization by a magnetic field, see e.g. books \cite{ref12}, reviews \cite{ref13}. The inverse effect will be considered elsewhere. At present, ME effects have attracted great interest for lots of applications: sensors, information storage, anti-radar coating \cite{ref14} and others \cite{ref13}. The conventional ME effect arises from interaction of electric and magnetic subsystems in time-asymmetric ionic crystals or piezoelectric and magnetic subsystems in composite multiferroics \cite{ref15}.

From applied point of view, linear ME effect has special importance \cite{ref13, ref14}. But also it requires specific efforts from fundamental point of view because the magnetic field is a pseudo vector, and electric polarization is a vector. So the linear effect in conventional multiferroics is only possible for a special asymmetry, e.g in multi-sublattice insulators with rare-earth elements. As it is shown below, the spintronics provides linear ME effect with high accuracy due to transfer of conduction electrons, which are absent in dielectric multiferroics.

In most experiments (see e.g. Ref.~\cite{ref18,ref19}), the value of the ME effect is 10-100 $\mu C/m^2$ at a magnetic field of~3~$T$ or more, giant ME effect is~3600~$\mu C/m^2$ at~7~$T$ \cite{ref20} with the same 10-100~$\mu C/m^2$ at~2~$T$.

In this paper, the strong ME effect of quite a different nature is predicted. It is related to the spin-dependent redistribution of free electrons in inhomogeneous magnetic conductive structures embedded into dielectric matrix, forming a metamaterial, i.e. nanostructured composite material, cf. for example Ref.~\cite{ref16}. In contrast to conventional metamaterials the purpose of this one is to reach high mutual capacity of meta-atoms (``artificial atoms'') from their closeness.

In the beginning let us consider one meta-atom of the metamaterial. Then we will show how ME effect occurs and why it increases while these meta-atoms decreasing.

The said meta-atom is a standard object in spintronics -- a micro-sized conductor consisting of two parts: a non-magnetic ($\mathcal N$) and magnetic ($\mathcal M$). At achievable stationary magnetic fields, the energy $\Delta$ of giant Zeeman splitting of magnetized $\mathcal M$-part (electron volts in ferromagnetic metals) is much larger than one of $\mathcal{N}$-part Zeeman splitting. Therefore, we will neglect the latter. The equilibrium of isolated meta-atom is determined by the constancy of the charge carriers' total number and equality of the electrochemical potentials of the meta-atom's magnetic and non-magnetic parts.

In equilibrium, the chemical potential level $\mu_\mathcal{M}$ in $\mathcal M$-part is the same for electrons with different spins due to the spin relaxation. Since the electron DOS~$N(\varepsilon)$ depends on the energy, the chemical potential~$\mu_\mathcal{M}$ differs from the chemical potential~$\mu_0$ of the system without a magnetic field. Let us analyze~$\mu_\mathcal{M}$ dependence on the magnetization and magnetic field.

For simplicity, we assume that the temperature is much lower than Fermi energy and Zeeman splitting energy in $\mathcal M$-part such that the Fermi distribution is replaced by a step function. Change of $\mu_\mathcal{M}$ in a magnetic field is considered in \cite{ref17} and experimentally-confirmed in\cite{ref17-1}:
\begin{equation}
\frac{d\mu_\mathcal{M}}{d(g\mu_BH)} = \frac{N_{\downarrow}^{-1} - N_{\uparrow}^{-1}}{2\left(N_{\downarrow}^{-1} + N_{\uparrow}^{-1} - 4I \right)},
\label{eq:mcdonald}
\end{equation}
where $g$ is the electronic g-factor, $\mu_B$ is the Bohr magneton, $H$ is the external magnetic field; $N_\uparrow=N(\mu_0+\Delta/2)$  and  $N_\downarrow=N(\mu_0-\Delta/2)$ are the electron DOS of the corresponding spin subbands at the Fermi level in the rigid band approximation, $N(\varepsilon)$ is the electron DOS (per one spin) in the absence of the Zeeman splitting, $I$ refers to Stoner parameter characterizing the magnitude of the electron exchange interaction in $\mathcal M$-part. Here the Zeeman splitting is:
\begin{equation}
\Delta = 2 (IM+\mu_B H),
\label{eq:delta-gen}
\end{equation}
where $M$ is the magnetization. From here, the variations of $\mu_\mathcal{M}$ in paramagnetic and ferromagnetic materials differ significantly because paramagnetic magnetization is proportional to $H$, while ferromagnetic material has its own magnetization.  

More specifically, in the case of paramagnetic material:
\begin{equation}
\Delta_{para}\approx\frac{2}{1-\alpha}\mu_B H,
\label{eq:delta-para}
\end{equation}
where $\alpha=2IN<1$ is the dimensionless parameter of exchange enhancement of the $\mathcal M$-part magnetic susceptibility for the paramagnetic material. When $\alpha\ge  1$  $\mathcal M$-part is ferromagnetic (Stoner criterion \cite{ref21}), the expression (\ref{eq:delta-para}) is not applicable, and
\begin{equation}
\Delta_{ferro}\approx 2IM
\label{eq:delta-ferro}
\end{equation}
does not depend on the external magnetic field in the main approximation by $H$.

Using (\ref{eq:mcdonald})-(\ref{eq:delta-ferro}) we obtain the following expression for the chemical potential variation in the case of the ferromagnetic $\mathcal M$-part ($\alpha>1$):
\begin{equation}
\delta\mu_{ferro}(H) = \frac{g\mu_B H}{2}\frac{X}{\alpha - 1},
\label{eq:delta-mu-ferro}
\end{equation}
where $X=(N_\uparrow - N_\downarrow)(N_\uparrow + N_\downarrow)^{-1}$ refers to spin polarization of DOS at the Fermi level. And for the case of paramagnetic $\mathcal M$-part with exchange enhancement ($\alpha<1$), we obtain:
\begin{equation}
\delta\mu_{para}(H) = -\frac{g\mu_B^2 N'(\varepsilon_F)}{4(1-\alpha)^2 N(\varepsilon_F)}H^2,
\label{eq:delta-mu-para}
\end{equation}
where $N'(\varepsilon_F)$ is the first derivative of the DOS with respect to the energy at Fermi level. Thus, the chemical potential change for paramagnetic $\mathcal M$ is $\propto H^2$, while it is linear for ferromagnetic $\mathcal M$, both far from the ferromagnetic transition with $H$. Also it is nonlinear and huge ($\sim IM$) at crossing the transition with $H$.

Under equilibrium condition, the electrochemical potentials of $\mathcal M$ and $\mathcal N$ parts of the meta-atom are equal, as has been said, for times longer than electron spin relaxation time. Because of non zero $\delta \mu_\mathcal{M}$, the electrochemical potentials equilibrate by the transfer of electrons between the $\mathcal N$ and $\mathcal M$ parts (transfer direction is determined by the sign of $\delta \mu_\mathcal{M}$).  
Now let's see what happens in a metamaterial which consists of these meta-atoms.

For simplicity, we consider a chain of identically oriented $\mathcal M$-$\mathcal N$ meta-atoms, Fig.1(a), separated by the insulator $\mathcal I$. The results for such a quasi-one-dimensional system is extended to three-dimensional systems.

\begin{figure}
\includegraphics[width=8 cm]{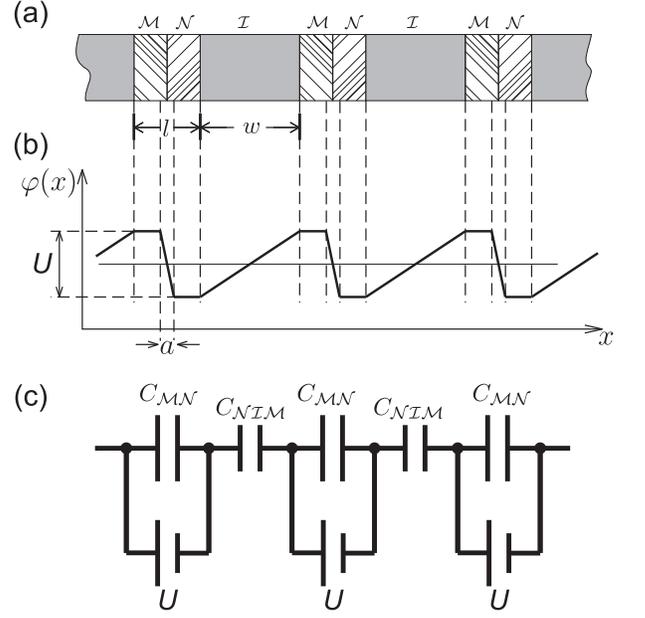}
\caption{(a)~Metamaterial is represented by~$\mathcal M$-$\mathcal N$-$\mathcal I$ chain, 
(b)~electrical potential for~$\delta \mu < 0$, 
(c)~equivalent circuit
} \label{fig1}
\end{figure}

As noted above, the appearance of $\delta \mu_\mathcal{M}$ induces the redistribution of charge carriers and subsequently the formation of an electric potential difference between $\mathcal M$ and $\mathcal N$ pieces \cite{refnote1}. 
This effect can be regarded as ``magneto-contact potential difference'', i.e. a part of the contact potential difference\cite{ref22} that depends on magnetic field. So the electric charges appear at the interfaces $\mathcal M$/$\mathcal N$, $\mathcal M$/$\mathcal I$, $\mathcal N$/$\mathcal I$. Thus the chain can be represented as an equivalent electrical circuit shown in Fig.1(c). In this figure, the $\mathcal M$-$\mathcal N$-$\mathcal I$ chain is a series of the two alternating capacities: $C_\mathcal{MN}$  is the capacity of $\mathcal M$/$\mathcal N$ interface and $C_\mathcal{NIM}$  is the capacity of $\mathcal N$/$\mathcal I$/$\mathcal M$ interface. The difference in electron chemical potential from change of a magnetic field is represented as power sources with the voltage of $U=e^{-1}\delta\mu$  connected to each of $\mathcal M$-$\mathcal N$ capacitances. 

It can be seen from the Fig.1(c) that the extra charges are concentrated at the $\mathcal M$/$\mathcal N$ and $\mathcal N$/$\mathcal I$/$\mathcal M$ interfaces. They are proportional to the capacity of the interfaces: $q_{\mathcal{MN}}=UC_{\mathcal{MN}}$,   $q_{\mathcal{NIM}}=UC_{\mathcal{NIM}}$ because of the potential equality for the interconnected plates of $C_{\mathcal{MN}}$ and $C_{\mathcal{NIM}}$. 
The electric potential as function of coordinate is shown at Fig.1(b). 
Using the formula for a flat capacitor, we get: $C_{\mathcal{NIM}}=\varepsilon S/(4\pi w)$, $C_{\mathcal{MN}}=S/(4\pi a)$, where $\varepsilon$ is the permittivity of $\mathcal I$, and $S$ is a cross-section, $w$ is the distance between $\mathcal M$-$\mathcal N$ meta-atoms, $a$ is a distance of charge separation on $\mathcal M$/$\mathcal N$ interface (of the order of screening length). The process of spatial separation of a charge results in appearance of the electric polarization\cite{refdip}:
\begin{equation}
\mathbf{P} = \frac{1}{V_{cell}}\int_{-\infty}^{+\infty} dt \int_{V_{cell}} dV \ \mathbf{j},
\label{eq:P-chain}
\end{equation}
where $V_{cell}$ is the volume of elementary cell and $\mathbf{j}$ is a density of current in charge redistribution and insulator polarization processes. Thus, using (\ref{eq:P-chain}) we obtain that the external magnetic field induces an electric polarization:
\begin{equation}
P(H)=\delta\mu(H)\frac{\gamma(\varepsilon-1)}{2\pi e(w+l)}.
\label{eq:P-result}
\end{equation}
So this is the ME effect. Here $0<\gamma<1$ is dimensionless coefficient of the packing density of the chains in the dielectric matrix. As can be seen, the direction of the electric polarization vector depends only on the relative position of $\mathcal M$/$\mathcal N$ structures in it, but not on the direction of the applied magnetic field. Also, as can be seen from (\ref{eq:P-result}) polarization will increase when dielectric permittivity of $\mathcal I$ increases, the size of $\mathcal M$/$\mathcal N$ structures decreases, and when they are located more compactly. Additionally, the expression (\ref{eq:P-result}) does not include parameters of material $\mathcal N$. This is due to the fact that the calculation does not take into account changes of the Fermi level by filling the electron quantum states. This correction is necessary only for extremely small meta-atoms (of the order of several interatomic distances). Note that it is the permittivity $\varepsilon \neq 1$ that makes nonzero electric polarization. Otherwise only net quadrupole will be created. So, insulator $\varepsilon \neq 1$ breaks the symmetry in charge redistribution, also it distinctly contributes to the total dipole moment.

The linear ME effect calculated by (\ref{eq:P-result}) and (\ref{eq:delta-mu-ferro}) is shown on Fig.3 for different thicknesses of $\mathcal N$/$\mathcal M$/$\mathcal I$ layers and the following parameters: $\mathcal M$-layer is Fe, $\mathcal N$-layer is Cu, insulator -- barium titanate (BaTiO$_3$)\cite{ref23} with $\varepsilon=1400$; we set $\gamma=0.5$, single multilayer thickness $l+w=10^{-6}$ cm for dashed line, $3\times 10^{-6}$ cm for solid line, $10^{-5}$ cm for dotted line. As the graphs shows, the magnitude of $P$ for submicron structures comes up to tens of $\mu C/m^2$ that is comparable with results Ref.~\cite{ref18, ref19}, and is obtained in quite a low magnetic field. The nanostructures (about 10 nanometers) are even better and the corresponding polarization is about hundreds of $\mu C/m^2$ at the same magnetic field. It is obvious from (\ref{eq:P-result}) that the $P$ can be increased by using the dielectric with a greater $\varepsilon$, such as the copolymers, ceramics, etc., or nonstationary nonintrinsic colossal $\varepsilon$, see e.g. \cite{ref19a} and references in it. To achieve the record even for $\varepsilon \gtrsim 1$, one need $\mathcal M$ so that the $H$ would cross the critical magnetic field point.

\begin{figure}
\includegraphics[width=6.3 cm]{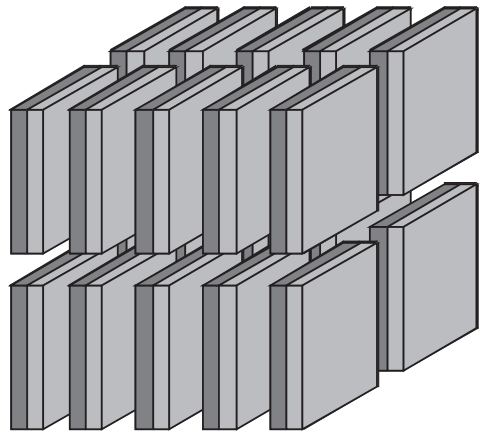}
\caption{The metamaterial consisting of a two-layer plates magnet/non-magnet in a dielectric matrix.} \label{fig2}
\end{figure} 
 
\begin{figure}
\includegraphics[width=8.2 cm]{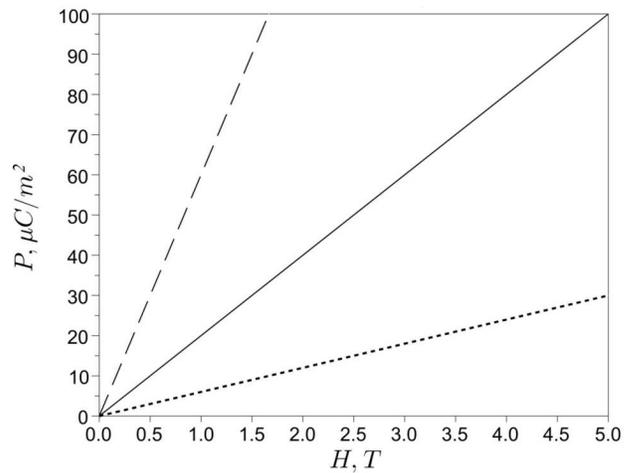}
\caption{The metamaterial Cu/Fe/BaTiO$_3$ electric polarization dependence on the external magnetic field. Dashed line: $l+w=10^{-6}$ cm  ; solid line: $l+w=3\times 10^{-6}$ cm ; dotted line: $l+w=10^{-5}$ cm} \label{fig3}
\end{figure}

If the $\mathcal M$-layer is paramagnetic with a strong exchange gain (e.g. Pd), the dependence of the electric polarization $H$, according to (\ref{eq:delta-mu-para}), (\ref{eq:P-result}) is quadratic and can be enhanced using $\mathcal M$-s which are close to satisfaction of the Stoner criterion of ferromagnetism and have a strong DOS dependence on the energy near Fermi level:
\begin{equation*}
\left.\frac{dN(\varepsilon)}{d\varepsilon}\right|_{\varepsilon=\varepsilon_F}>>\frac{N(\varepsilon_F)}{\varepsilon_F}.
\end{equation*}
 
In summary, we have substantiated that metamaterial consisting of oriented isolated inhomogeneous magnetic conductors can exhibit strong (regarding known range of ME effect) electric polarization response on a magnetic field. It is interesting that the polarization direction is determined by location and orientation of the $\mathcal N$-$\mathcal M$ structures in the material. This feature allows to separate predicted effect even from other ME effects that can be imposed, in particular, if the intrinsic ME coupling is present in insulator $\mathcal I$. In such a case we propose changing the sample orientation relative to the external magnetic field in the measurements.

It has been shown that linear ME effect occurs when $\mathcal M$-part of meta-atoms is ferromagnetic. As indicated above, the linear ME effect is interesting in that $H$ is a pseudo vector, and $P$ is a vector. In our effect $P$ is composed of the true scalar (average energy of the electron magnetic moment in magnetic field) and true vector (the direction of electron transfer during the equilibration). The accuracy of the linear behavior is very high because it's determined by $\delta \mu/\varepsilon_F \sim 10^{-3} \div 10^{-4}$.

In the case of the paramagnetic $\mathcal M$-parts, ME effect is quadratic in $H$. In particular, it can be detected by a frequency-doubled response to a variable magnetic field. Here we have applied equilibrium approach. It is valid for quasi-static magnetic field when the characteristic frequency is less than the reverse time of the nuclear and electron spin relaxation ($10^9\div 10^{12}$ sec$^{-1}$). The results of the study in variable electromagnetic fields of higher frequency will be published later.

Notice one more application of the effect. Since the magnetized state is preserved after ferromagnetic transition, the electric polarization of the  metamaterials can also retain for a long time. Therefore, these materials may be used as magnetically driven electrets, i.e. substances with long-lasting electrically polarized state.

Finally, we emphasize that the electric dipole moment of standalone meta-atom is extremely small, because the charge separation is confined only within a very thin layer in the vicinity of the $\mathcal M$-$\mathcal N$ interface because of absence of neighbors' mutual capacity. Therefore, only the nanostructured metamaterial exhibits strong ME response.

This work was supported by the grant of the NAS of Ukraine \#4/13-N. The authors thank Prof. A. I. Kopeliovich for continuing interest in this work and valuable comments.


\begin{thebibliography}{99}
\bibitem{refnote1} Of course each $\mathcal M$ or $\mathcal N$ piece is equipotential.
\bibitem{ref23} It is well known: there is no ME effect in ferroelectric BaTiO$_3$. So the ME contribution can be easily separated from the residual polarization.
\end{thebibliography}
\end{document}